\documentclass[twocolumn,prstab,floatfix]{revtex4}
\usepackage{graphicx}
\usepackage{amsmath}
\usepackage{latexsym}
\usepackage{amsfonts}
\usepackage{amssymb}
\usepackage{subfigure}
\usepackage{psfrag}

\begin{document}
\renewcommand{\vec}{\mathbf}
\newcommand{\vhat}[1]{\mathbf{\hat{#1}}}
\newcommand{\diffsum}[1][]{\ensuremath{\Delta_{#1}/\Sigma}}
\newcommand{\poisson}{{\sl POISSON}}

\title{Orbit and Optics Improvement by Evaluating the Nonlinear BPM Response
in CESR}
\author{Richard W.~Helms} 
\author{Georg H.~Hoffstaetter}
\affiliation{Laboratory for Elementary Particle Physics,
Cornell University, Ithaca, New York 14853}

\begin{abstract}
We present an improved system for orbit and betatron phase measurement
utilizing nonlinear models of BPM pickup response.  We first describe the
calculation of the BPM pickup signals as nonlinear functions of beam position
using Green's reciprocity theorem with a two-dimensional formalism.  We then
describe the incorporation of these calculations in our beam position
measurements by inverting the nonlinear functions, giving us beam position as
a function of the pickup signals, and how this is also used to improve our
measurement of the betatron phase advance.  Measurements are presented
comparing this system with the linearized pickup response used historically at
CESR.
\end{abstract}

\maketitle
\section{Introduction}
CESR measures beam position and betatron phase with approximately one hundred
beam position monitors (BPMs) distributed around the storage ring.  Each BPM
consists of four button-type electrodes mounted flush with, and electrically
isolated from, the surface of the beam pipe.  A moving particle bunch induces
charge on the beam pipe walls and on the surface of each button, which one can
describe as image currents or as surface charge due to the transverse component
of the bunch's electric field~\cite{shafer}.

The BPM buttons are connected to electronics that process and record signals
which are a function of the distance between the button and the passing bunch.
The four signals from each BPM are used to determine the beam
position and betatron phase advance.  At many accelerators, the button signals'
nonlinear dependence on the beam position is linearized for simplicity.  Before
the improvements described here, this approach was also used in CESR.

Our efforts to improve the beam position measurements by including the
nonlinear BPM response is motivated by CESR's pretzel orbits, where electron
and positron beams avoid parasitic collisions by following separate paths with
large displacements from the central axis of the beam pipe.  The linearized
methods are not reliable for such large amplitudes, and have made accurate beam
position and betatron phase measurements at CESR impossible under colliding
beam conditions.  We will illustrate those shortcomings and present
measurements demonstrating improvement by using the nonlinear models.

\section{Background}
\begin{figure}[htb]
  \psfrag{S1}{$S_1$}
  \psfrag{S2}{$S_2$}
  \psfrag{S3}{$S_3$}
  \psfrag{S4}{$S_4$}
  \psfrag{x}{$x$}
  \psfrag{y}{$y$}
  \centering\includegraphics[width=2.5in]{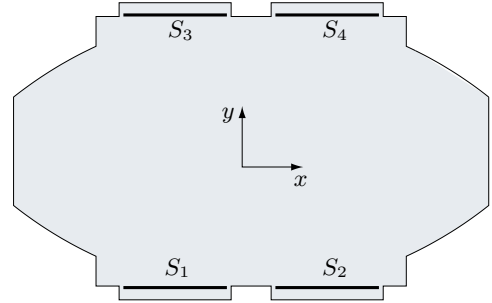}
  \caption{Arrangement of buttons in CESR arc BPMs}
  \label{fig:buttons}
\end{figure}
Many accelerators, including CESR, have traditionally assumed a linear
relationship between the beam position and the BPM button signals.  Given four
signals $S_i\:(i=1,\ldots,4)$ from buttons arranged as in
Fig.~\ref{fig:buttons}, the transverse beam position is given approximately by
\begin{eqnarray}
   x & = & k_x \frac{(S_2 + S_4) - (S_1 + S_3)}{\Sigma_i S_i}
       \label{eqn:diffsum_x} \\
   y & = & k_y \frac{(S_3 + S_4) - (S_1 + S_2)}{\Sigma_i S_i}
       \label{eqn:diffsum_y}
\end{eqnarray}
where $k_{x,y}$ are scale factors set by the geometry of each BPM type.
This evaluation of BPM signals is often called the {\it difference-over-sum}
method.  Equations~(\ref{eqn:diffsum_x}-\ref{eqn:diffsum_y}) provide an
estimation of the bunch position in relatively few arithmetic operations.

Because analytical approaches to determining the factors $k_{x,y}$ make
drastic approximations to the BPM geometry, we have tried to measure the
factors experimentally at CESR through a variety of techniques summarized in 
Table~\ref{tab:scalefactors}.  Those include translating a section of the beam
pipe containing the BPM with precision actuators, using a test stand with a
movable antenna to simulate the beam, and using the known value of the
dispersion while changing the beam energy in dispersive
regions~\cite{bagley:beampipe}.
\begin{table}[htb]
  \centering
  \begin{tabular*}{3in}{|l|@{\extracolsep{\fill}}r|r|} \hline \hline
    Method & $k_x$ (mm) & $k_y$ (mm) \\ \hline
    20~MHz antenna                    & 25.58$\pm$.33 & 20.58$\pm$.43 \\ 
    Dispersion (1990)                 & 26.3          & \\ 
    Dispersion (1991)                 & 27.4$\pm$.6   & \\
    Moving beam pipe ($e^+$)          & 26.82$\pm$.25 & 19.96$\pm$.11 \\
    Moving beam pipe ($e^-$)          & 27.14$\pm$.54 & 20.48$\pm$.19\\
    2D {\sl Poisson} model            & 26.2          & 19.6 \\ \hline\hline
  \end{tabular*}
  \caption{Measured scale factors for CESR arc BPMs.}
  \label{tab:scalefactors}
\end{table}

But precise knowledge of $k_{x,y}$ is of limited benefit, since 
Eqs.~(\ref{eqn:diffsum_x}-\ref{eqn:diffsum_y}) yield only the linear part of
the signal dependence for bunches near the center of the BPM.  In the next
section, we describe our technique for accurately calculating button signals,
but let us first use those results to illustrate the limitation of the
linearized formulae.
 
\begin{figure}[htb]
  \centering\includegraphics{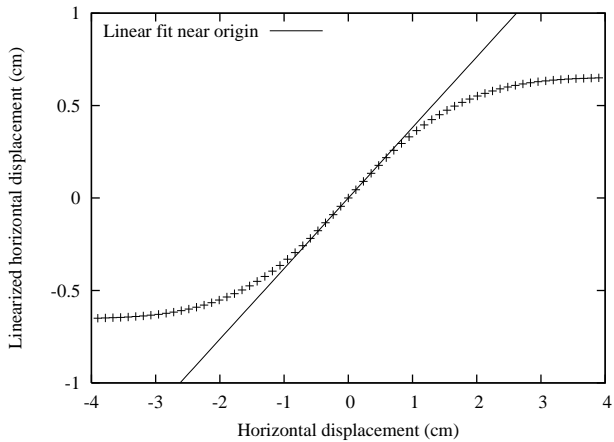}
  \caption{Linearized horizontal position measurement in an arc BPM.  At a
    typical pretzel amplitude of 1.5~cm, the linearized formula shows
    significant disagreement with a linear fit.}
  \label{fig:diffsum}
\end{figure}
In Fig.~\ref{fig:diffsum}, the four button signals were calculated numerically
for a beam at different horizontal displacements.  The four signals were
combined according to Eq.~(\ref{eqn:diffsum_x}) and plotted against the known
horizontal displacement.  The slope near the origin gives $k_x^{-1}$, but the
linear relationship breaks down noticeably at approximately 1~cm, and beyond
2~cm the relationship fails completely.  Because pretzel orbits in CESR are
typically as large as 1.5~cm, this is precisely what has hindered accurate
measurements under colliding beam conditions until the improvements described
in this paper were implemented.

\begin{figure}
  \centering\includegraphics{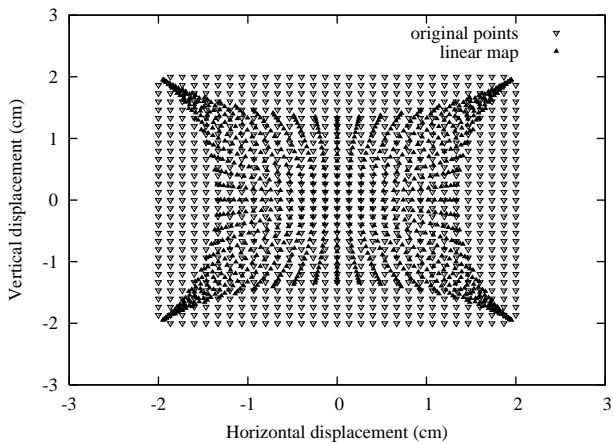}
  \caption{Linearized map distortion in CESR interaction-point BPM with
    approximately circular cross-section.}
  \label{fig:diffsumgrid_ip}
\end{figure}
\begin{figure}
  \centering\includegraphics{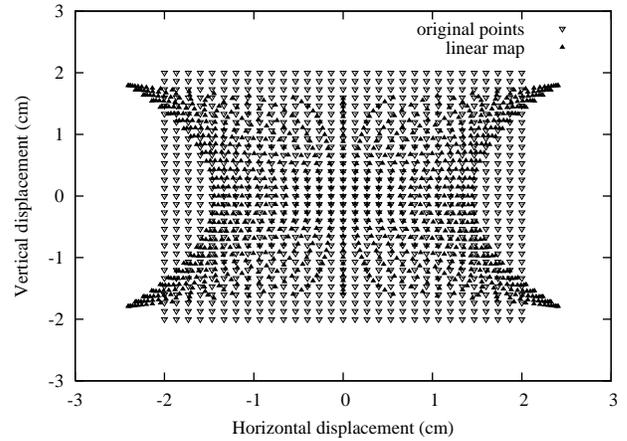}
  \caption{Linearized map distortion in CESR arc BPM with approximately
    elliptical cross-section.}
  \label{fig:diffsumgrid_arc}
\end{figure}

The problem of this nonlinearity is made even more evident in two dimensions.
Figures~\ref{fig:diffsumgrid_ip}-\ref{fig:diffsumgrid_arc} show a regular grid
of $(x,y)$ points and the mapping of those same points under
Eq.~(\ref{eqn:diffsum_x}-\ref{eqn:diffsum_y}).  Both BPMs show the
characteristic {\it pincushion} distortion, which increases with distance
from the origin.

In CESR, betatron phase measurements also rely on a related assumption about
the linearity of the button signals.  The betatron phase is measured by shaking
the beam at a sideband of the betatron frequency.  For each detector, the phase
for each button is calculated by electronically comparing AC signal on that
button to the phase of the shaking.  From the individual horizontal (vertical)
button phases $\theta_{h,i}\:(\theta_{v,i})$, the horizontal and vertical
betatron phase is calculated by
\begin{eqnarray}
  \label{eqn:oldphase_x}
  A_h e^{i\theta_h} & = & e^{i\theta_{2,h}} + e^{i\theta_{4,h}} 
       - e^{i\theta_{1,h}} - e^{i\theta_{3,h}} \\
  \label{eqn:oldphase_y}
  A_v e^{i\theta_v} & = & e^{i\theta_{3,v}} + e^{i\theta_{4,v}}
       - e^{i\theta_{1,v}} - e^{i\theta_{2,v}},
\end{eqnarray}
where $A_{h,v}$ are real constants that are not used
further~\cite{sagan:old_phase}.  Other than the minus signs which account for
the assumption that the beam is shaking between the pairs of buttons, this is
simply an averaging of button phases represented as complex vectors.

When the horizontal orbit amplitude is large, the beam begins to shake
{\it underneath} the buttons, and the relationship between the beam motion
and the button signal becomes complicated.  In such cases, some of the buttons
may report an inaccurate phase, and averaging them with the rest corrupts the
final answer.  We will show how our nonlinear models can improve not only beam
position measurements, but these measurements as well.

\section{An Improved System For Position And Phase Measurement}
In order to overcome the limitations described, a new system has
been implemented with two major components: realistic numerical models of the
button response, and an efficient algorithm for inverting the model to yield
beam position.

\subsection{Numerical Calculation of BPM Response}
For accurate beam position measurements, a function is required that expresses
the bunch location $(x,y)$ as a nonlinear function of the button signals.
Since the four button signals lead to two coordinates (and a scale factor),
the problem is over-constrained, and this function cannot be obtained
directly.  The inverse (button signals from beam position), however, is readily
obtainable by standard numerical techniques.

The most accurate and direct method of numerical solution is to simulate the
bunch in a three-dimensional BPM, calculating the electromagnetic fields, and
from them, the charge on the buttons.  The simulation could be repeated for
different beam locations, and the fields recalculated until enough solutions
were accumulated to describe the behavior over the entire BPM.  However,
this is very computationally intensive, and can be avoided by the methods that
follow.

\subsubsection{Two-Dimensional Approximation}
For ultra-relativistic bunches in a beam pipe with constant cross
section, the approximate electromagnetic fields can be calculated
using a two~dimensional formalism~\cite{cuperus,krinsky}. Assuming the
bunch has negligible transverse extent, the charge distribution of the
bunch may be written, in the lab frame, as
\begin{equation}
  \rho = \delta(\vec{r-r_0}) \sum_k \rho_k \cos\left(k(z-vt)\right)
\end{equation}
where the longitudinal dependence in $z$ has been written as a Fourier
expansion.  Transforming to the reference frame of the bunch, the charge
density and electric potential are written
\begin{eqnarray}
  \rho^* &=& \delta(\vec{r-r_0})
     \sum_k\frac{\rho_k}{\gamma}\cos(kz^*/\gamma), \\
  \Phi^* &=& \Phi(\vec{r})\sum_k\frac{\phi_k}{\gamma}\cos(kz^*/\gamma).
\end{eqnarray}
We write Poisson's equation $\nabla^2\Phi^*=\rho^*$ in the bunch frame as
\begin{equation}
  \left(\nabla^2_\perp - \frac{k^2}{\gamma^2}\right) \Phi(\vec{r})\phi_k =
      \delta(\vec{r-r_0})\rho_k
\end{equation}
where $\nabla^2_\perp$ is the two-dimensional transverse Laplacian.
For bunches with length $\sigma_l$ without appreciable longitudinal
substructure, $\rho_k$ is only relevant for
$k\leq\frac{1}{\sigma_l}$. The characteristic distance over which
$\Phi(\vec r)$ changes is the diameter $a$ of the beam-pipe so that the
order of magnitude estimate $|\nabla^2_\perp\Phi(\vec
r)|\approx\frac{1}{a^2}|\Phi|$ can be made. For sufficiently long
bunches and sufficiently large values of $\gamma$, the relevant values
of $k/\gamma$ can be neglected, i.e. when
$\frac{1}{\gamma^2}\ll(\frac{\sigma_l}{a})^2$ and the solution is
described by the two~dimensional, electrostatic case
\begin{equation}
  \nabla^2_\perp \Phi(\vec{r})=\frac{\rho_k}{\phi_k}\delta(\vec{r-r_0}).
\end{equation}
Since we only need $\Phi(\vec{r})$ up to a multiplicative factor, we don't
worry about the constant coefficients on the right-hand side.

\subsubsection{Green's Reciprocity Theorem}
Rather than perform a separate calculation of the button signals for many beam
positions, we use Green's reciprocity theorem to calculate the button signals
for all $(x,y)$ inside the BPM with a single numerical calculation.
This theorem states that the surface charge $\sigma$ on a button due to a
test charge at $(x,y)$ is proportional to the potential at that same position
when the test charge is absent and the button is excited by a potential
$\mathcal{V}$.

Suppose we have two scalar functions $\phi_1$ and $\phi_2$ in a volume $V$
bounded by a surface $S$.  We form the vector field
\begin{equation}
  \vec{A}=\phi_1\nabla\phi_2
\end{equation}
for which the divergence theorem guarantees
\begin{equation}
  \label{eqn:div_thm}
  \int_V\nabla\cdot\vec{A}\,dV = \oint_S\vec{A}\cdot\vhat{n}\,da.
\end{equation}
Manipulating the integrands gives
\begin{eqnarray}
  \nabla\cdot(\phi_1\nabla\phi_2) & = & (\nabla\phi_1)\cdot(\nabla\phi_2)
                                    + \phi_1 \nabla^2 \phi_2 \\
  \vec{A}\cdot\vhat{n} & = & \phi_1\nabla\phi_2\cdot\vhat{n} =
                             \phi_1 \frac{\partial\phi_2}{\partial n}
\end{eqnarray}
where $\vhat{n}$ is a unit vector normal to the surface and pointing out of
the volume of integration, and $\partial/\partial n$ indicates differentiation
with respect that direction. Equation~(\ref{eqn:div_thm}) yields
\begin{equation}
  \label{eqn:bigintegral}
  \int_V\left[(\nabla\phi_1)\cdot(\nabla\phi_2) + 
    \phi_1 \nabla^2 \phi_2\right]\,dV =
  \oint_S\phi_1 \frac{\partial\phi_2}{\partial n}\,da.
\end{equation}
If we interchange $\phi_1$ and $\phi_2$ and
subtract the result from Eq.~(\ref{eqn:bigintegral}), we can eliminate the
first term in the integrand of the left hand side.  This gives
\begin{equation}
  \int_V\left[\phi_1\nabla^2\phi_2-\phi_2\nabla^2\phi_1\right]\,dV =
  \oint_S\left[\phi_1 \frac{\partial\phi_2}{\partial n} -
    \phi_2 \frac{\partial\phi_1}{\partial n}\right]\,da.
\end{equation}

Taking the $\phi_i$ to be potentials for volume charge density $\rho_i$ and
surface charge density $\sigma_i$ leads to \emph{Green's reciprocity theorem}:
\begin{equation}
  \label{eqn:greensreciprocity}
  \int_V \phi_1 \rho_2 \,dV + \oint_S \phi_1 \sigma_2 \, da =
  \int_V \phi_2 \rho_1 \,dV + \oint_S \phi_2 \sigma_1 \, da
\end{equation}
where we have used $\nabla^2\phi=-\rho$ and $\partial\phi/\partial n=\sigma$
(recall that $\vhat{n}$ points \emph{into} the conducting surface).

Connecting this result to the case of a BPM, imagine $\phi_1$ corresponds to
the potential when a single button is excited with a potential $\mathcal{V}$
and all other surfaces are grounded.  We can calculate the potential
$\phi_1(x,y)$ by numerical solution of Laplace's equation.
For the second potential $\phi_2$, we ground all surfaces and put a charge
distribution $\rho_2(x,y)$ inside the BPM.

We plug the two cases into Eq.~(\ref{eqn:greensreciprocity}) and observe that
the third integral vanishes because there is no volume charge for the first
case ($\rho_1=0$ in $V$).  The fourth integral vanishes because we grounded the
beam pipe and the buttons ($\phi_2=0$ on $S$).  Since $\mathcal{V}$ can be
pulled out of the second integral, what remains is just the total charge on the
button, labeled $q_b$, giving
\begin{equation}  \label{eqn:green_result}
  \int_V \phi_1(x,y)\rho_2(x,y)\,dV = -\mathcal{V} q_b.
\end{equation}

If $\rho_2$ is a point charge $q$ located at $(x_0,y_0)$, then the integral in
Eq.~(\ref{eqn:green_result}) picks out the value $\phi_1(x_0,y_0)$.  We arrive
at the final relation
\begin{equation}
  q_b = -\frac{q\phi(x_0,y_0)}{\mathcal{V}},
\end{equation}
remembering that $\phi(x,y)$ and $\mathcal{V}$ refer to the two different
configurations.

Therefore, since the signal on a button is proportional to the induced surface
charge on that button $q_b$, $\phi(x_0,y_0)$ is the solution to the problem of
calculating the button signal, up to a multiplicative constant, as a function
of the bunch location.

\subsubsection{Calculation of the Button Signals}
Based on the previous arguments, we use \poisson\ to solve the boundary value
problem for $\phi(x,y)$.  For the two-dimensional boundary, we take a slice at
the longitudinal midplane of each BPM.  The first button is set (arbitrarily)
at 10~volts and all other surfaces are grounded.  \poisson\ generates a mesh
inside the boundary, computes the solution to Laplace's equation on the mesh,
and stores the result at regular grid points in an output file.

CESR BPMs have multiple geometric symmetries, so the signals $\phi_i(x,y)$ on
the other three buttons are just reflections or rotations of the coordinates
for the excited button in the first calculation of $\phi_1(x,y)$.  To compute
$\phi_i(x,y)$ between grid points, we use bicubic interpolating polynomials,
which are stored for quick subsequent evaluation.

\subsection{Realtime Inversion}
For beam position measurements, we start with button signals $S_i$ and seek the
location $(x,y)$ of the beam.  The result $\phi_i(x,y)$ from the \poisson\
calculation must be inverted, and
since we have four constraints (four buttons) and three parameters (position
$(x,y)$, and a scale factor) we proceed by {\it fitting} the calculated button
signals to the measured signals.  We minimize the merit function
\begin{equation}
  \label{eqn:chisq}
  \chi^2 = \sum_{i=1}^4\frac{\left(q\phi_i(x,y)-S_i\right)^2}{\sigma_i^2},
\end{equation}
where $\phi_i(x,y)$ is the signal on the $i^\mathrm{th}$ button and the
$\sigma_i$ are the uncertainties in the measured signals (which we take to be
the same for all four buttons).  The factor $q$ is proportional to the beam
current and could be used for beam loss studies.

Minimization is performed via the Levenberg-Marquardt method provided in
Numerical Recipes.  This requires an initial guess for the parameters, which we
find by scanning only the grid points of $\phi(x,y)$ (without evaluating the
interpolating polynomials) for the values of the parameters that minimize
$\chi^2$.  Then we iteratively minimize over the continuous functions,
typically arriving within less than $10^{-6}$~m of the minimum after
six steps.

\subsection{Phase Measurements}
We can improve our measurement of the betatron phase advance between BPMs by
incorporating our knowledge of the nonlinear button response.  In this
measurement, the beam is excited to small oscillations around its equilibrium
position $(x_0,y_0)$.
Let the phase and amplitude of the AC signal on the $i^\mathrm{th}$ button be
represented by the complex number $\vec{C}_i$, and let the phase and amplitude
of the horizontal and vertical components of the oscillatory beam motion be
represented by complex numbers $\vec{A}_x$ and $\vec{A}_y$, respectively.  To
first order, their relationship is given by
\begin{equation}\label{eqn:newphase}
  \vec{C}_i = r_{i,x}\vec{A}_x + r_{i,y}\vec{A}_y,
\end{equation}
where the $r_{i,(x,y)}$ are given by
\begin{eqnarray}
  r_{i,x} & = & \left.q\frac{d\phi_i(x,y)}{dx}\right|_{(x_0,y_0)} \\
  r_{i,y} & = & \left.q\frac{d\phi_i(x,y)}{dy}\right|_{(x_0,y_0)}.
\end{eqnarray}
The $\phi_i$ are the functions described in the previous section.  Their
derivatives are easily calculated from the coefficients of their interpolating
polynomials.

Given the measured $\vec{C}_i$, we calculate $\vec{A}_x$ and $\vec{A}_y$ by
minimizing
\begin{equation}\label{eqn:phase_chi2}
  \chi^2 = \sum_{i=1}^4 \frac{1}{\sigma_i^2}|r_{i,x}\vec{A}_x + 
  r_{i,y}\vec{A}_y-\vec{C}_i|^2.
\end{equation}
Since the $\sigma_i$ depend on the closed orbit deviation and the values of
$\vec{A}_{x,y}$, the minimization must also be performed iteratively.  The
horizontal and vertical phase advance is then given by the complex phase of
$\vec{A}_x$ and $\vec{A}_y$.  Whenever a horizontal excitation creates a
vertical amplitude, or vice versa, this method is used in CESR to compute the
coupling coefficients also.

\section{Results}
Testing the new system presents a challenge in that we can only produce
controlled large amplitude orbits with the electrostatic separators.  Since
the separators are calibrated from BPM measurements, they do not provide an
independent check on our ability to measure large amplitudes accurately.  Our
strategy, therefore, must be to use other measurements to check the accuracy
at small amplitudes, and then confirm the expected linear relation between
the separator strength and the beam position at large amplitudes.

\begin{figure}[htb]
  \centering
  \includegraphics{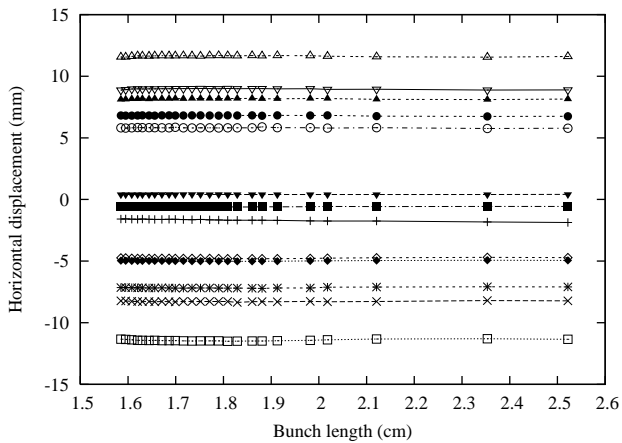}
  \caption{Beam position at various detectors showing little or no bunch
    length dependence.}
  \label{fig:bunch_length}
\end{figure}
To perform a two-dimensional approximation, we argued that the bunches are
sufficiently long.  To verify that assumption, we have looked experimentally
for a bunch length dependence in large amplitude orbits.  With the pretzel at
its nominal value of about 1.5~cm closed orbit deviation, the bunch length was
calculated from the measured synchrotron tune, which we adjust by changing the
RF accelerating voltage.  As Fig.~\ref{fig:bunch_length} illustrates, the beam
position shows little or no dependence over the range of bunch lengths we
expect in CESR.

\begin{figure}[htb]
  \centering
  \includegraphics{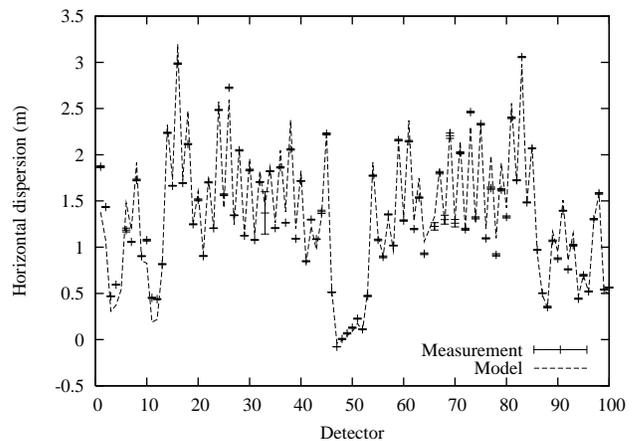}
  \caption{Measured and calculated dispersion.}
  \label{fig:eta}
\end{figure}
Changing the RF frequency in CESR changes the beam energy, and in dispersive
regions, changes the beam position by up to a few millimeters.  Measuring the
beam position at many different energies allows us to measure the
dispersion, which we compare to the theoretical value from the lattice in
Fig.~\ref{fig:eta}.  This agreement verifies the small amplitude, or linear
part of our nonlinear models.

\begin{figure}[htb]
  \centering
  \includegraphics{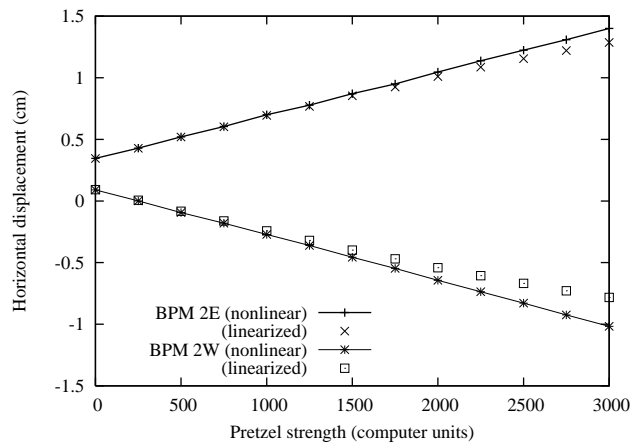}
  \caption{Beam position at two detectors calculated with the nonlinear and
    linearized methods.}
  \label{fig:pretzel_linearity}
\end{figure}
To observe the large amplitude accuracy of the new system, we rely on the
electrostatic separators to change the orbit amplitude linearly.  By
increasing the horizontal separator strength, we observe in
Fig.~\ref{fig:pretzel_linearity} that the orbit calculated with the nonlinear
method does show the correct behavior, while the orbit calculated with the 
linearized formula shows the deviation that was illustrated in
Fog.~\ref{fig:diffsum}.

\begin{figure}[htb]
  \centering
  \includegraphics{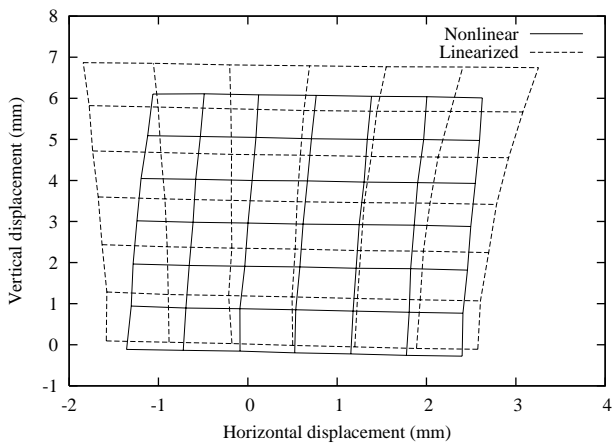}
  \caption{Separator scan.  Orbits at detector 9W calculated using linearized
    (dashed) and  nonlinear (solid) methods.}
  \label{fig:pretzscan}
\end{figure}
To demonstrate improvement in two dimensions, the voltages on individual
horizontal and vertical separators were scanned over a regular grid.  The
measured beam positions should also lie on a regular grid, which is shown in
Fig.~\ref{fig:pretzscan}.  Some sheering is evident in the plot, which
may be due to coupling of the vertical and horizontal motion between the
separator and the BPM, or to a rotation of the
BPM.  The pincushion effect is notably reduced with the new calculation.

\begin{figure}[htb]
  \centering
  \includegraphics{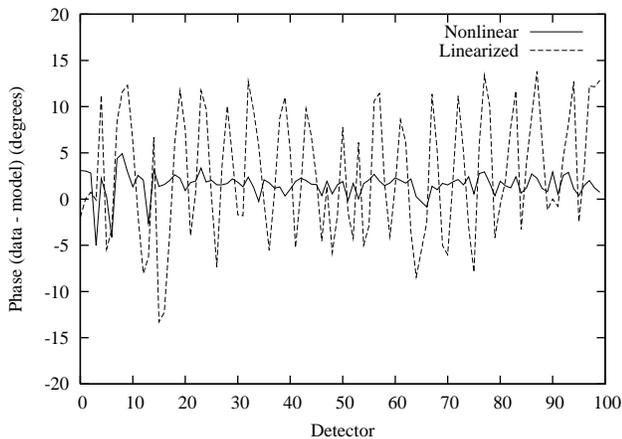}
  \caption{Difference in horizontal betatron phase advance between data and
    model with large closed orbit distortion after using a phase correction
    algorithm based on the linear (dashed) and nonlinear (solid) BPM
    evaluation.}
  \label{fig:phase}
\end{figure}
We use betatron phase measurements to correct the difference between the
physical optics and the values in our model lattice.  Without the nonlinear
correction, large closed orbit distortions hindered this process since the
data we sought to fit did not correspond to the actual phase.
Figure~\ref{fig:phase} shows the drastically improved agreement we can achieve
between the model phase and the data when the new BPM calibration is used.

\section{BPM Calibration}
The response of a particular BPM may differ from that of the computational
model (linear or nonlinear) for a variety of reasons.  The leading candidate
for this effect is the variation in insertion depth of the individual buttons
(i.e., the distance from the button surface to the surface of its cylindrical
housing).  This manifests itself as different gains for the signals from
different buttons.

Following the method of~\cite{lambertson,keil:thesis}, we determine the gain
for each button from the capacitive coupling between each pair of buttons.  If
the ideal coupling between two buttons is given by $U_{ij}$, then the measured
coupling will be $\tilde{U}_{ij}=b_i b_j U_{ij}$ where $b_i,b_j$ are the
effective gains of the input and output button, respectively.  Symmetric pairs
of buttons have equal ideal coupling, so $U_{12}=U_{34}$, $U_{13}=U_{24}$,
and $U_{14}=U_{23}$.

Using this symmetry for the six measurements
$\tilde{U}_{ij}\:(i=1,\ldots,3,\:j=i+1,\ldots,4)$
we can calculate the four $b_i$ up to an multiplicative factor.  Normalizing to
$b_1$ gives
\begin{eqnarray}
b_1 &=& 1\ ,\label{b1}\\
b_2 &=&
\sqrt{\frac{ \tilde{U}_{23}\tilde{U}_{24}}{\tilde{U}_{13}\tilde{U}_{14}}}\ ,\\
b_3 &=&
\sqrt{\frac{ \tilde{U}_{23}\tilde{U}_{43}}{\tilde{U}_{12}\tilde{U}_{14}}}\ ,\\
b_4 &=&
\sqrt{\frac{ \tilde{U}_{24}\tilde{U}_{43}}{\tilde{U}_{12}\tilde{U}_{13}}}\ .
\end{eqnarray}

These gain coefficients are used to correct the button signals before
calculating the beam position.

\begin{figure}[htb]
  \centering
  \includegraphics{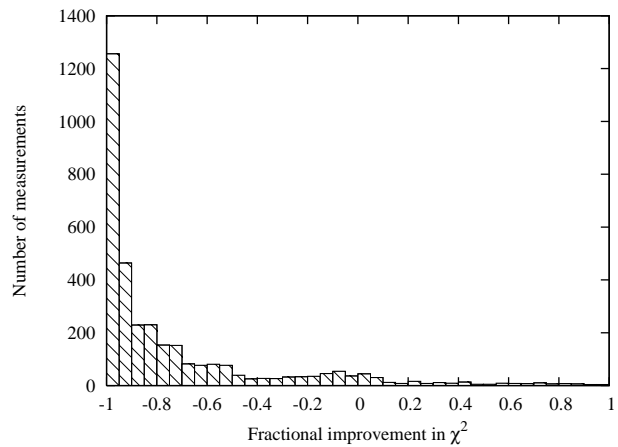}
  \caption{Fractional improvement in the $\chi^2$ of the beam position fit
    due to the calibration coefficients ($-1=100\%$ improvement).
    }
  \label{fig:chi2_improvement}
\end{figure}

\begin{figure}[htb]
  \centering
  \includegraphics{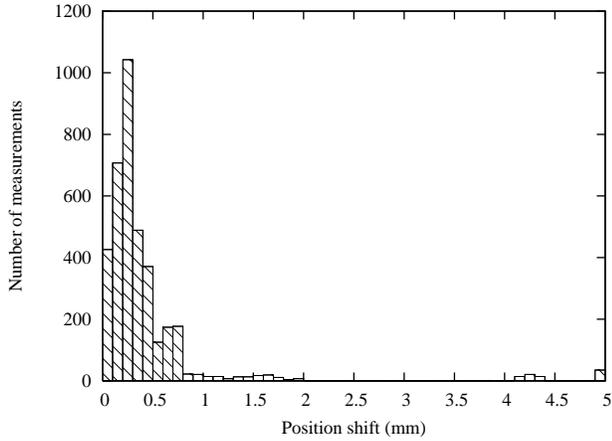}
  \caption{Correction in the calculated position due to the calibration
    coefficients.}
  \label{fig:position_shift}
\end{figure}

With data drawn from approximately 3700 individual beam position measurements,
Fig.~\ref{fig:chi2_improvement} shows that when these coefficients are
employed, the $\chi^2$ of the fit between the measured signals and the modeled
signals is significantly reduced.  The resulting correction to the calculated
position is shown in Fig.~\ref{fig:position_shift} to be approximately 0.5~mm.

\section{Conclusion}
Two-dimensional, electrostatic models of BPM pickup response have been
used with great success at CESR to measure beam position and betatron phase
advance for large closed orbit distortions.

Calibration of our BPMs has reduced measurement errors due to button
misalignments.

\section{Acknowledgments}
The authors wish to thank REU student Beau Meredith for his calibration
of most of CESR's BPMs.

\bibliography{bpm}
\end{document}